\documentclass [preprint]{aastex}
\usepackage{graphicx}

\begin{document}

\title{ANALYSIS OF A STATE CHANGING SUPERSOFT X-RAY SOURCE IN M31}

\author{B. Patel\altaffilmark{1}, R. Di Stefano\altaffilmark{2},
  T. Nelson\altaffilmark{3}, F. A. Primini\altaffilmark{2},
  J. Liu\altaffilmark{2}, S. Scoles\altaffilmark{2}}
\altaffiltext{1}{Department of Physics and Astronomy, Rutgers, The State University of New Jersey, Piscataway, NJ 08854-8019, USA}
\altaffiltext{2}{Harvard-Smithsonian Center for Astrophysics, 60 Garden st. Cambridge, MA 02138, USA}
\altaffiltext{3}{Department of Physics, 1000 Hilltop Circle, University of Maryland at Baltimore, Baltimore, MD 21250, USA}
\begin{abstract}
We report on observations of a luminous supersoft X-ray source (SSS) in M31, r1-25, that has exhibited
spectral changes to harder X-ray states. We document these spectral
changes. In addition, we show that they have important implications
for modeling the source. Quasisoft states in a source that has been
observed as an SSS represent a newly-discovered phenomenon. We show
how such state changers could prove to be examples of unusual black
hole or neutron star accretors. Future observations of this and other state changers can provide the information needed to determine the nature(s) of these intriguing new sources.

\end{abstract}


\section{INTRODUCTION}

Luminous supersoft X-ray sources (SSSs) were established as a class by
{\it ROSAT} observations of roughly 30 sources in the
 Magellanic Clouds, Milky Way, and M31 \citep{Gre}. {\it Chandra} and {\it XMM-Newton} observations of external galaxies have now discovered hundreds of soft X-ray sources with properties that both exemplify and extend the class of SSSs \citep[e.g.,][]{Dis2003, Kong2003, Dist2003, DisK2004, Gre2004, Dis2006, Dis2010, Orio2010, Liu}. Even though they are bright, with luminosities higher than $10^{36}$~erg~s$^{-1}$, we know of only  a handful in the Galaxy, because the radiation they emit is readily absorbed by the interstellar medium. In fact, the SSSs used to define the class display little or no emission above $1$~keV. Roughly a dozen SSSs are known in the Magellanic Clouds \citep{Gre}. Some of these are associated with novae, and are clearly hot white dwarfs \citep[see][]{Gre}. In M31, some SSSs have been shown to be associated with supernova remnants and novae \citep{Orio2010, Piet2005, Piet2007, Stiele2010}.

The most mysterious component of the class is comprised of X-ray binaries,
most with orbital periods of a day or less. A promising model for
these sources is one in which their prodigious luminosities are
produced by the nuclear burning of matter accreted by a white dwarf
\citep{Van, Rap}. Nuclear burning should allow the
white dwarf to retain accreted matter and increase in mass. Binary
SSSs have therefore been suggested as progenitors of accretion-induced
collapse \citep{Van} and of Type~Ia supernovae \citep{Rap}. Nuclear-burning
white dwarfs in wider orbits are also expected \citep{Hach, Dis1996}. Indeed, symbiotic
binaries have been observed as SSSs \citep[e.g.,][]{Gre, Orio2007}.

Because it is difficult to detect SSSs in the Milky Way, and the
Magellanic Clouds are too small to host a large population, it is
important to search for SSSs in external galaxies. The advent of {\it
  Chandra} and {\it XMM-Newton} has made such searches
possible. Hundreds of SSSs have now been discovered, some in galaxies
as far from us as the Virgo cluster \citep{Liu}.
As the numbers of SSSs has increased, we have begun to find evidence
of sources that have properties different from those of the 
SSSs that established the class. Some SSSs are hundreds of times more
luminous than the Eddington limit for a Chandrasekhar-mass white
dwarf. Several of these ultraluminous supersoft sources are candidates
for accreting black holes \citep{Dis2004, Kong2004, Kong2005, Mukai2005, Liu2008}. In addition, the
search for the softest sources has identified a class of sources that are
significantly harder than SSSs, yet also significantly softer than
canonical X-ray sources: quasisoft X-ray sources \citep[QSS;][]{Dist2004, DisK2003, Dis2004}. QSSs have luminosities
above $10^{36}$~ergs~s$^{-1},$ but emit few or no photons with energy
above $2$ keV. Some could be hot white dwarfs
in which there is an additional hard component and/or which are highly
absorbed. Those fitting this model are good candidates for progenitors
of Type~Ia supernovae, because they would likely correspond to the
most massive nuclear-burning white dwarfs \citep[e.g.,][references]{Rap, Dis2010a}. Others are too hot to be white dwarfs and may  correspond to either black holes or neutron stars.

In this paper we report on a source (r1-25) that has been observed to
switch between SSS and QSS states \citep{Stiele2008, Stiele2010, Dis2010b, Orio2010}. Unlike M101-ULX-1 \citep[e.g.,][]{Kong2004, Kong2005, Mukai2005}, a well known state changing source, r1-25 is not ultraluminous.  r1-25 is unique in that no such sources are known in the Galaxy or Magellanic Clouds. We examined all
available {\it Chandra},  {\it Swift}, and optical data from Hubble Space Telescope (HST) and the 
Local Group Survey (LGS) for this source. We also checked the literature for {\it
  XMM-Newton} observations and analysis of the source. The question we want to
answer is: what is the physical nature of this unique source?

We analyze the X-ray data for the source in $\S$2. In $\S$3, we present evidence that the source changes state. In $\S$4, we analyze the optical data. We discuss the possible models that fit this source in $\S$5.

\section{ X-ray Observations and Analysis}

The source r1-25 has been discussed in the literature \citep{Kong2002, Will, Dis2004, Kaaret, Voss, Stiele2010, Dis2010b, Orio2010}. It is located in the central region of M31 approximately $0.4\arcmin$ (about 91 parsecs) from the
nucleus. The coordinates of the source are (J2000.0) RA = 00:42:47.90, DEC = +41:15:49.99. This region of M31 has been well sampled over the past 13 years. For this paper, we searched the {\it Chandra} archive for all public observations of r1-25 through June 2, 2009. There were 86 observations that
covered the source. r1-25 was detected in 45 observations (28 ACIS-I, 2
ACIS-S, and 15 HRC-I observations) from August 8, 2000 to
March 11, 2009. We note that the source position was the same in all
detections, and we are confident that r1-25 is one source. We supplemented the {\it Chandra} data with {\it Swift}
observations of the region in 2009. However, the source was off during
the {\it Swift} observations, which are not included in this paper. 

\citet{Liu} presented the photometry for the ACIS observations of r1-25 taken between 2000 and 2004. We analyzed the ACIS data taken between 2004 and 2009 using the method presented in that paper. We used CIAO version 4.1.2 \citep{Frus} to
analyze the recent data. For source detection, we used CIAO
tool {\tt wavdetect} \citep{Free, Frus}. Only photons between 0.3-7 keV were considered and the following bands
were used to classify them: Soft, 0.3-1.1 keV; Medium, 1.1-2 keV; and
Hard, 2-7 keV. We note that there were no photons detected below 0.3 keV. The total count rate, as well as the rate in each band, were corrected by a vignetting factor. The vignetting factor was derived from the exposure map as the ratio between the local and the maximum map values.     

For HRC-I observations, we used the CIAO tool {\tt dmextract}
\citep{Frus} to extract raw counts in source and background
apertures, as shown in Figure 1, and to compute net counts. We defined
aperture sizes so that the source aperture encircled energy fraction
was $\sim$1 and the background aperture fraction 0. We also used the CIAO
{\tt aprates} \citep{Frus} tool to compute the
background-marginalized posterior probability distribution for the
source rates, assuming non-informative prior distributions for source
and background rates \citep{Kas}. We note that the HRC data
were not corrected for vignetting. Thus, the rates are lower than the
ACIS counterparts by $10\%$ or less. For observations with low flux significance, we used the posterior probability distributions to estimate the 3$\sigma$ upper limits to the rate. The upper limits were calculated using {\tt aprates} and the same method was applied to both HRC and ACIS observations.

Table 1 shows the photometry results for r1-25.  Columns 4, 5, and 6 show the Soft, Medium, and Hard count rates,
respectively. The uncertainties are 1$\sigma$ in size and calculated using Poisson statistics. Figure 2 shows the light curve for r1-25. The top panel shows the count rate in each band (soft, medium, and hard) vs. obsdate for the ACIS-I detections; it demonstrates that the relative count rates
change over time for the source. The bottom panel of Figure 2 shows the total count rate in the ACIS-I band vs. obsdate. We used the {\tt pimms} \citep{Mukai1993} to convert ACIS-S and HRC-I count rates to ACIS-I units. We choose ACIS-I units because most of the detections of the source were in this instrument. We indicate ACIS-S and HRC-I observations as a range of values, and assumed {\it N$_H$} = 1.1 $\times$ $10^{21}$ cm$^{-2}$ for the conversion. For OBSID 1854, we assumed a thermal blackbody model with temperatures of 75 and 83 eV  to determine the lower and upper limits of the range, respectively. For OBSID 1575, we used  $k\, T$ = 120 eV and $k\, T$ = 150 eV to determine the range with the same column. For HRC-I detections and upper limits, as well as ACIS-S upper limits, we converted to ACIS-I units using $k\, T$ = 75 eV and $k\, T$ = 300 eV.

\section{Evidence for State Change}

We used the CIAO tool {\it dmextract} to create source spectra for all
observations where more than 60 source counts were collected. Source spectra were extracted from a
circular region with 
radius 6 pixels centered on the source. Associated background spectra
were extracted from an annular region (also centered on the source)
with inner and outer radii of 10 and 25 pixels, respectively. For all
observations, we created spectral response files using the CIAO tasks
{\it mkacisrmf} and {\it mkarf} \citep{Frus}. 

We modeled the resulting spectra using the XSPEC package \citep{Arnaud}.  Since the number of source counts is small, we fit the spectra (binned to have one count per bin, as suggested in the XSPEC documentation)  using the C-statistic \citep{Cash} rather than the $\chi^{2}$ statistic, which has been shown to introduce a systematic bias in parameter estimation in low count rate spectra.  The C-statistic as implemented in Xspec models the source and background spectrum simultaneously, scaling the background spectrum channel by channel to the size of the source region \footnote{See http://heasarc.nasa.gov/xanadu/xspec/manual/XSappendixStatistics.html}. One $\sigma$ uncertainties were estimated using the error command.  We found that model fits to observations with less than 60 counts were not constrained; for this reason, we do not include them in our spectral analysis of r1-25.  The four Obs-IDs with more than 60 counts were: 1575 (ACIS-S, 183 source counts), 4720 (ACIS-I, 61 source counts), 4721 (ACIS-I, 65 source counts) and 4722 (ACIS-I, 62 source counts).

We fit a simple absorbed blackbody model to each spectrum in the
energy range 0.3-8.0 keV, using the wabs model for the interstellar
absorption \citep{Morr}. None of the spectra had enough
counts below 1 keV to place a tight constraint on the interstellar
column density (in several cases the fits were consistent with no
absorption), so we fixed {\it N$_H$} at two values representative of
the range expected towards an X-ray source in M31 (1.1 $\times$
10$^{21}$ and 6.4 $\times$ 10$^{21}$ cm$^{-2}$). The lower limit was
taken from \citet{Dis2004}. The upper limit is
shown for completeness; it is very unlikely that the source has such a
large column density. The resulting parameter values and their 1
$\sigma$ uncertainties are presented in Table 2. We note that our
results for ObsID 1575 (we found 0.110 keV $\le$ $k\, T$ $\le$ 0.130 keV)
are consistent with those reported in \citet{Dis2004} (they found that $k\, T$ = 0.122 keV). Examining the temperatures found for r1-25, a clear increase in the later observations can be seen, independent of the choice of absorbing column. Furthermore, r1-25 significantly changes in luminosity when it changes state. At the 90\% confidence level, the temperature found for the ACIS-S spectrum is $\sim$0.1 keV lower than the ACIS-I values. The blackbody temperature returned by the model fitting is constrained primarily by the high energy cut off. Although ACIS-I has poorer low energy sensitivity, the fact that significantly harder counts are detected in such short exposures indicates that the increase in model temperature is real and independent of the differences between the two ACIS instruments.

\section{Optical Observations and Analysis}

The location of r1-25 has been observed with the ACS
camera onboard \textit{HST}. The source has been observed 4 times: on 2004-01-23 (observation j8vp03010) for 2200 seconds, 2004-08-15 (observation j8vp05010) for 2200 seconds, 2006-02-10 (observation j9ju01010) for 4360 seconds, and 2007-01-10 ( observation j9ju06010) for  4672 seconds. Images are only available in the F435W filter (approximately equal to
the \textit{B} filter in ground based systems). Only one {\it Chandra}
observation (ObsID 8183) was taken within a week of an ACS observation
(j9ju06010) of the source. However, the source was not detected in
ObsID 8183. 

All data were obtained using the Wide Field Channel (WFC), which has a
$202\arcsec$ $\times$ $202\arcsec$ field of view \citep{May}. Each
observation was carried out in the standard four pointing dither
pattern \citep{May}. These individual images were then combined using the PyRAF task MultiDrizzle \citep{Fruc}, which also removes cosmic rays and corrects the geometrical distortion which results from the orientation of the ACS with respect to the HST focal plane.  We chose not to apply an automatic background subtraction in order to perform photometry, since there is a steep gradient in the diffuse light this close to the center of M31 making background estimation unreliable.  Finally, we utilized the ability of MultiDrizzle to resample the spatial scale of the image, resulting in a final pixel scale of $0.025\arcsec$/pix.  

The World Coordinate System (WCS) in HST images can be offset from
standard reference frames by as much as one arcsecond \citep{May}. To improve the astrometry, we registered the final
drizzled images to the WCS of the Local Group Survey (LGS) images of
M31 \citep{Mass}. Stars common to both images were
identified, and their centroid positions calculated using the IRAF
task imcentroid. We then used the task ccmap in IRAF to update the WCS
of the HST images.  The final rms (1$\sigma$) errors in the alignment
were of order 
0.006$\arcsec$ in RA, and 0.002$\arcsec$ in declination. We note that the
RMS errors on the alignment of the HST images to the WCS of the LGS
survey were always smaller than 0.01$\arcsec$ (which is smaller than
one pixel in the rescaled images).  

We also aligned the deepest {\it Chandra} observation (ObsID 1575)
with the WCS of the LGS images using the same procedure applied to
the HST images. We found an alignment error of 0.109$\arcsec$ in RA and
0.149$\arcsec$ in DEC. The centroid position in the corrected
WCS for the source is RA:00:42:47.90, DEC:+41:15:49.99, with errors of
0.08$\arcsec$ (RA) and 0.07$\arcsec$ (DEC). To get the final positional
error, we added the alignment and centroid errors in quadrature. The
final RA error is 0.13$\arcsec$ and final DEC error is
0.17$\arcsec$ (these are 1$\sigma$ errors). 

Since the WFC images cover such a large area and contain so many
stars, we extracted subimages 1000 pixels on a side centered on our
source of interest and performed photometry on those.  We used the
DAOPHOT II and ALLSTAR routines \citep{Stetson} to find and
photometer stars in the images.  Stars suitable for calculating a
point spread functions (PSFs) were identified by hand to avoid
problems due to crowding.  The final measured counts for each star
were converted first to count rate by dividing by the exposure time of
each observation, and then to AB system magnitudes using the
conversion factors in the ACS users handbook \citep{May}.

We show the final reduced image for the source in Figure 3,
superimposed with the X-ray 3$\sigma$ positional error
ellipse. The ellipse is drawn using the 3$\sigma$ RA error
(0.39$\arcsec$) as the semi-major axis and the 3$\sigma$ DEC error
(0.51$\arcsec$) as the semi-minor axis. The location of r1-25 is
extremely crowded, and being so close to the core is 
subject to a high background of diffuse light.  In our series of four
images, two 
(j9ju01010 and j9ju06010) are
also significantly deeper, which affects the completeness of the stars we
can detect.  The small
size of the Chandra error circle does however simplify the analysis of
the photometry, since very few sources are inside the X-ray
3$\sigma$ error ellipse.
Examining the images, a number of sources are detected inside the
error circle, although most are unresolved. We note that there are no
catalog stars within the {\it Chandra} error ellipse. In fact, only one source
is resolved inside the error ellipse in all four images by the
photometry source detection algorithm.  We have marked this source
with a red circle in Figure 3.   

The single resolved source in Figure 3 has observed magnitudes of 24.45
$\pm$ 0.06, 24.27 $\pm$ 0.05, 24.31 $\pm$ 0.04 and 24.41 $\pm$ 0.04 in
each of the four images, where the uncertainties are 1$\sigma$ in size. Using the standard $E(B-V) = 0.062$ for M31 \citep{Sch}, the extinction in the F435W filter is 0.785
magnitudes. Thus, the star's $M_B$ is between -0.80 and -0.98 in the four
observations, assuming $(m-M)_0$ = 24.47 \citep{Holl}. This
demonstrates that, within the uncertainties, there is no evidence of
variability in this object. Although other sources are picked up by the DAOphot detection algorithm inside the error ellipse in some images, these additional detections can be accounted for by the longer exposure time, or are unreliable due to crowding.

\citet{Grupe} looked at the spectral energy distribution of 92
active galactic nuclei that had soft X-ray spectra. The AGN they
studied had comparable X-ray count rates to r1-25. However, the AGN
were much brighter in the B band (14 $<$ $m_{B}$ $<$ 18 ) than any source
in Figure 3. For this reason, we are confident that r1-25 is not an AGN. 

With no color information, we cannot determine what the object marked
in Figure 3 is for certain. If it is a star, it would correspond to a
late B type with bolometric luminosity of $~10^{36}$ ergs s$^{-1}$. We
note that it is too luminous in the B band to 
be a red giant, and too dim to be a red supergiant.

\section{Models}

The state changes observed in r1-25 are extremely unusual for an SSS. In this section, we consider a number of physical
models with the goal of uncovering the nature of the X-ray emitting
source.  In order for any model of r1-25 to be successful, it must be
able to explain all of the observed features of the source. These
features include the source's appearance as an SSS-HR source in
observation 1575 with ACIS-S, with kT $\sim$ 130 eV and a 0.3--8 keV
luminosity of 4 $\times$ 10$^{36}$ erg s$^{-1}$. We wish to emphasize
once again that even though r1-25 had an effective temperature in
excess of 100 eV when detected as an SSS 
(much higher than most SSSs), it satisfied the strictest SSS criterion
as defined in \citet{Dis2004}. That is, the detection in
observation 1575 (when the source was an SSS) had no hard
counts, medium counts consistent with zero, and at least 3$\sigma$ detection
in the soft band. 

Furthermore, the model must explain the subsequent detections of the
source as a $\sim$250 eV source, \textit{with higher luminosity} than
in the soft state ($\sim$10$^{37}$ erg s$^{-1}$). Lastly, the model
must be consistent with an optical 
counterpart with F435W magnitude fainter than $\sim$24.3 ($M_B$ fainter than
$\sim$ -1). We note that white dwarf, neutron star, and black hole
SSSs are consistent with this optical constraint.  The following subsections outline white dwarf, neutron star and black
hole models, and compare their features to the observed properties of
r1-25. 

\subsection{White Dwarfs} 

White dwarfs that have recently experienced novae have temperatures
and luminosities that can make them detectable as SSSs.\footnote{A
  specific post-nova system will be detectable as an SSS only if the
  white dwarf stays hot enough to be emitting as an SSS until after
  the optical depth has decreased enough to let radiation escape \citep[e.g.,][]{Sala2005}.} Many
SSSs detected in M31 were recent novae \citep[e.g.,][references therein]{Piet2005, Piet2007, Stiele2010}. When the
white dwarf cools, some novae are 
detected as harder X-ray sources, but the X-ray luminosity is around
$10^{32}-10^{33}$~ergs~s$^{-1}$ \citep{Sala2010}. In contrast, novae
in a supersoft state are detected with $k\, T$ $\sim$ 50 eV and L$_X$ of
$10^{37}-10^{38}$~ergs~s$^{-1}$ \cite[e.g.,][references therein]{Stiele2010}. Thus r1-25, even in its softest state (with kT $\sim$130 eV
and L$_X$ of 4 $\times$ 10$^{36}$ erg s$^{-1}$), is too hard to be
consistent with the very soft emission detected in typical supersoft
novae. Also, the source is not 
consistent with the harder states of novae. That is r1-25 in its hardest
state ($\sim$250 eV and L$_X$ of 1.1 $\times$ 10$^{37}$ erg s$^{-1}$)
is softer and more luminous than novae in their hard states. Moreover, unlike novae, the harder states of r1-25 are more luminous than its soft state. We therefore
turn to models in which a white dwarf accretes matter at high rates.

When a white dwarf accretes mass at a high enough
rate that the incoming matter can experience nuclear burning, the
white dwarf can appear as an SSS. In this case, the source will not be
in a hard state. The copious energy we receive from
such sources is provided
by nuclear burning, rather than accretion. The effective radii are
comparable to the white dwarf radii, so the emission can be characterized by
values of $k\, T$ in the range of tens of eV for low-mass white dwarfs,
and $\sim100$ eV for white dwarfs approaching the Chandrasekhar
mass ($M_C$). For each white dwarf mass, 
nuclear burning can occur only within a narrow range of accretion
rates \citep[][references therein]{Nom, Iben, Fuj, Shen}. These rates are very high:
$\sim10^{-7}M_\odot$ for a solar-mass white dwarf, ten times higher for a
white dwarf with mass near $M_C$ \citep{Dis2010a}. 
At such rates of infall, accretion alone produces luminosities in the
range of $10^{36}-10^{37}$~ergs~s$^{-1}$, typically a few percent of
the total energy of the system.

Consider a case, when its accretion rate places a white dwarf near the
lower end 
of the steady-burning region, or just below it. In this case,
nuclear burning may be episodic. During and just after nuclear-burning
episodes, the emission is dominated by soft emission. As the white
dwarf cools, however, it becomes  less luminous and the emission is
dominated by accretion. Although at high rates of accretion, the 
emission is expected to be softer than typical for low-accretion-rate
white dwarfs, such as cataclysmic variables, it can nevertheless
be harder than typical of SSSs \citep[e.g.,][]{Pop, Pat}. Thus, the source could appear to be
quasisoft or hard. If the donor is a giant or a Roche-lobe filling
star in a circular orbit,
the accretion rate should could continue to be high. The source
will continue to be detected as a harder source. Nuclear burning
episodes occurring at intervals ranging from months to decades would
make the source more luminous and detectable as an SSS. 

The points made above about the quasi-steady nuclear burning white dwarf
model are illustrated in Figure 4. The figure is a plot of $k\, T$ vs. the
logarithm of bolometric luminosity, LOG[L] for 
various quasi-steady nuclear burning white dwarfs (green points) along
with the r1-25 spectra (black and blue points). The points represent the XSPEC spectral fits of r1-25 shown in Table 2. The
solid black points represent the fits where we assume L$_X$ =
L. The X-ray luminosity, however, is not equivalent to the bolometric
luminosity. We assume that the X-ray luminosity represents, at least
a quarter of the bolometric luminosity. Thus, we plot the
open blue circles which represent XSPEC fits assuming L$_X$ =
$0.25L$. The plot clearly indicates that the quasi-steady nuclear
burning white dwarf model does not fit r1-25, as the data are too hard
and/or dim to fit the model. That is, none of the XSPEC points fall in the range of quasi-steady nuclear burning white dwarf.

\subsection{Neutron Stars} 

\def\ns{neutron star} 
\def\lu{luminosit}
\def\mf{magnetic field}
\def\acc{accretion} \def\accg{accreting} 
Isolated neutron stars and neutron stars in quiescent 
low-mass X-ray binaries (qLMXBs) have been observed with spectra
in the SSS or QSS range, but they are typically 3-5 orders of
magnitude less luminous than the ``classical'' SSSs and QSSs 
\citep[e.g.,][]{Hab, Pires}. 
Neutron stars accreting at high rates, however,
 have luminosities in the range (above $10^{36}$~erg~s$^{-1}$)
observed for SSSs and QSSs. But, at the time SSSs were discovered,
all known \accg\ \ns s emitted hard x-rays. The lack of hard
emission from SSSs therefore  
seemed to be more easily accommodated in white dwarf models.

Nevertheless,  
\citet{Kyla} showed
that accreting 
\ns s can be observed as SSSs under the right circumstances.
They considered near-Eddington accretion through a disk. 
Radiation pressure from the inner disk 
can push some plasma into an ``extensive outer disk corona.'' 
If the corona extends to large
enough radii, and if it is optically thick, the neutron star will radiate as an SSS.
Indeed, there is observational support for the 
idea that neutron stars can produce very soft spectra.
For example, \citet{Hughes} discovered a transient pulsar in the 
Small Magellanic
Cloud that has an unpulsed, highly luminous (near Eddington)
soft $\sim 35$~eV component. If such a system were to be viewed from an
angle at which the hard radiation is not detected, it  
would have the properties associated with the ``classical'' SSSs
first discovered in the Magellanic Clouds \citep{Long}. 

Although the details of the model considered by 
\citet{Kyla} may not apply to all systems, their work,
combined with observations, indicates that \ns\ models  
must be considered. Yet, beyond the success of the white-dwarf
models, there is another reason that neutron star models have not been
popular, and that was alluded to by \citet{Kyla}. This can
be simply stated by saying that the physics determining the size of the 
photosphere was put in by hand. Thus, the solutions for radial flows that
extend out to {\sl at least} a few thousand \ns\ radii produce SSS-like
behavior, and radial flows that extend out to {\sl at most} a few hundred
\ns\ radii produce more standard LMXB-like behavior. The range of 
photospheric radii between these two extremes would be associated
with luminous emitters of thermal radiation with $k\, T$ in the
range between roughly $100$~eV and a few hundred eV; 
that is, the sources would be
QSSs, which had not yet been discovered. 

Here we point out that 
the discovery of QSSs provides reasons to revisit neutron star
models, eliminating the need for fine tuning problem that 
\citet{Kyla} encountered. 
The key issue to address is what determines the size of the 
photosphere. It is likely to be linked to accretion rate, with
higher rates capable of producing larger photospheres, as in
the \citet{Kyla} model. Here we suggest that in some cases, 
the edge of the magnetosphere could roughly correspond to the 
photosphere. If this is the case, then, given that the 
mass, radius, and magnetic field of the accretor are
all roughly constant over time scales of months to years, the
photospheric radius would be driven primarily by changes in
the accretion rate. Equation (1) and Figure 5 show the relationship
that would be predicted between the temperature and luminosity.   

Consider a neutron star producing 
near-Eddington soft emission.
If this emission emanates from a nearly spherical photosphere
with radius equal to the Alfven radius, then
\begin {equation} 
k\, T = 100.4\, {\rm eV}\, 
\Bigg(\frac{10^{11}{\rm G}}{B_s}\Bigg)^{\frac{2}{7}}   
\Bigg(\frac{10\, {\rm km}}{R_{NS}}\Bigg)^{\frac{5}{7}}   
\Bigg(\frac{M_\odot}{M}\Bigg)^{\frac{1}{14}}   
\Bigg(\frac{L}{10^{37} {\rm ergs\, s}^{-1}}\Bigg)^{\frac{11}{28}}   
\end {equation} 
In this expression, $R_{NS}$ is the radius of the neutron star,
$B_s$ is the value of magnetic field on the surface, and $M$ and $L$ 
represent the neutron star's mass and luminosity, respectively. 
We derive this expression from equation 11 
in \citet{Ost}.\footnote{The form we show in equation 1 uses 
the equation for the mass accretion rate, 
$\dot{M}_{acc}=\frac{L_{acc}R_{NS}}{GM}$, and 
the blackbody luminosity equation, $L=4\pi\sigma(r_a)^2 T^4$. In 
these equations, r$_a$ is the Alfven radius and T is 
the blackbody temperature.} 
An interesting feature of this expression is that 
there are ranges of
reasonable values of the physical parameters in which the $k\, T$ is in
the range expected for SSSs or QSSs.
Furthermore, for a specific neutron star, the effective temperature
depends on L which can change as the accretion rate changes.
Since changes in accretion rate are common, we may therefore expect
the effective temperatures of some neutron stars to change. Depending
on the physical parameters, these changes could produce transitions
from SSS to QSS states.

Figure 5 shows two plots of $k\, T$ vs. LOG[L]. The top panel shows
several curves which differ from each other in the value of $B_s$, which
changes by 
a factor of ten between curves, as shown. The bottom plot shows 
the r1-25 spectra
along with the $B_s=10^{10}$~G curve. The black points 
represent the XSPEC fits shown in Table 2. The plots were made in 
the same manner as in Figure 4 (see $\S$5.1), except that we assumed
neutron star models have L $\approx$ L$_X$. Figure 5 shows that 
the distribution of points
is roughly consistent with what is
expected for the neutron star model discussed above. The
XSPEC points seem to follow the same trend as the curve, with some
variation. 

The general agreement between the trend of increasing $k\, T$ with
increasing luminosity is promising, and  
neutron-star models in
which the photospheric radius is not governed by the size of the 
Alfven radius may follow similar trends, with $k\, T$ increasing with
luminosity. 
Nevertheless, 
this general 
trend is not unique to  
neutron star models, as we will see in \S 5.3, where black hole models
are considered. It is therefore  
important to develop observational criteria that can
identify the nature of the compact accretor. 

For example, the neutron-star natures of LMXBs in
the Galaxy's globular clusters have been verified through
the detections of both bursts \citep[e.g.,][]{Lewin}
and pulsed \citep[e.g.,][]{Zhel} radiation. Both types of
variable components are expected to be harder than the dominant
softer radiation from an extended photosphere. 
Thus, for example in an x-ray pulsar,  
the diagnostic for the neutron star model would be periodicity in the 
arrival times of the harder photons.  
If, therefore, we can identify QSSs and state changers
in the Milky Way or in the Magellanic Clouds, we can test models
by searching for evidence
of hard bursts or pulses.
Detecting these in state changers, or in QSSs, 
would be possible if the system is close enough, and 
would verify the neutron-star nature of the accretor.

\subsection{Black Holes}

Accreting black holes can exhibit thermal-dominant states in which
the emission is dominated by a component emanating from the inner
portion of the accretion disk \citep[e.g.,][]{Rem}. SSSs and QSSs have
therefore both been suggested as possible black holes. In fact, the
most well-known state changer is M101-ULX-1, an ultraluminous SSS that
has been detected also in high QSS and low-hard states
\citep[e.g.,][]{Kong2004, Kong2005, Mukai2005}. M101-ULX-1 is almost
certainly a black hole. Its mass could be either in the range typical
of Galactic stellar-mass black holes or else in the higher range
($50-10^4\, M_\odot$) suggested for intermediate-mass black holes. 

The luminosity of r1-25 is 1-3 orders of magnitude 
smaller than the luminosities measured for M101-ULS-1
when it is in a soft state. It is therefore highly unlikely
to be an intermediate-mass black hole. In fact, if the luminosity
is less than roughly a percent of the Eddington luminosity,
then the inner disk will not be optically thick and the emission will
not be thermal. This suggests that, if this source is a black hole, 
it is more likely to be of stellar mass. The top panel of Figure 7,
first shown in \citet{Dis2010b},
shows that QSS emission is expected in the thermal-dominant
state of black
holes with mass below $\sim 100\, M_\odot.$ The radius of the inner
disk would determine the value of the effective temperatures; the
spectrum could be either QSS or SSS. At lower rates of accretion,
the emission would be hard. Note that there are two curves for each mass; the top curve assumes the
inner portion of the accretion disk is at 6MG/c$^2$ (3r$_s$, where r$_s$ is the Schwarzschild radius) and the
bottom assumes 18MG/c$^2$ (9r$_s$). Note that that the top curve (at 3r$_s$) represents the maximum luminosity and temperature for this
model in which the radiation comes from the inner disk.

Figure 6 shows the r1-25 spectra along with the
10$M_\odot$ and 100$M_\odot$ curves. The black points representing the XSPEC fits shown in Table 2 were plotted in the same manner as in Figure 5 (see $\S$5.2). The distribution of the points with XSPEC fits is roughly
consistent with what is expected for a black hole of roughly
10$M_\odot$. 

\section{Conclusion}

We have tracked the long-term behavior of the M31 X-ray source
r1-25. First observed by ROSAT on 1990-07-24, then by both {\it
  Chandra} and {\it XMM-Newton}and most recently by Swift, r1-25 is
one of the best-studies soft X-ray sources. There are 86 public {\it
  Chandra} observations of the source through June 2, 2009, with 45
detections. For {\it XMM-Newton}, there are 26 public observations of
the source, with \citet{Stiele2010} reporting detections in only the
2004 data. The detections of the source start in 1999 and continue
through 2009.  

By doing this we have documented the fact that r1-25 has transitioned
from an SSS to a harder, QSS state. In the SSS state its estimated
X-ray luminosity is a few times $10^{36}$ ergs $s^{-1}$, and the
luminosity appears to be higher, but not much over $10^{37}$ ergs
$s^{-1}$ in the harder state. Only one other X-ray source, M101-ULX-1,
has well-studied state changes \citep[e.g.,][]{Kong2004, Kong2005,
  Mukai2005}. While M101-ULX-1, which has been observed with X-ray 
luminosity as high as $10^{40}-10^{41}$ ergs 
$s^{-1}$, is almost certainly a black hole, the nature of r1-25 is
more difficult to establish, because its luminosity range is
consistent with white dwarf, neutron star, or black hole accretors. 

Whatever its nature, its behavior is different from anything we have observed. We
have shown that the observed behavior is consistent with a 
black hole accretor with a mass in the 10$M_\odot$ 
range. In this case, our observations of r1-25 have all found it to
be in a thermal-dominant state. The inner disk radius would have been
larger in the SSS state. If r1-25 is a black hole with a mass of
approximately $10M_\odot,$ it could be more luminous in future
observations, if the donor star is able to contribute mass at a higher
rate. Should the luminosity approach the Eddington luminosity, the
system would be unlikely to remain in the thermal dominant state, and
hard emission could be detected. Similarly, if the luminosity falls
below $\sim 1\%$ of the Eddington value, the spectrum would likely be
hard. 

We have also shown that the observed behavior of r1-25 is consistent with a
neutron star accretor with $B_s$ = $10^{10}$ G. We assume that the
magnetic field should be constant over the short interval of
observations of the source. The model suggests that the harder
states are more luminous than the softer ones, which is consistent
with the r1-25 spectra. We note that neutron star models are
testable if we can find state changers and QSSs in the Milky Way or
Magellanic Clouds, as both bursts and pulsed radiation would be
detectable in nearby neutron stars.

White dwarf models are the least likely fit for
r1-25. The source does not seem to exhibit behavior of a post-nova system,
as its spectrum is harder than novae that are SSSs (even when r1-25 is
in its soft state). The source is also too luminous and soft to be
consistent with novae in their harder states. Furthermore, quasi-steady
nuclear burning white dwarf models do not fit the data. We have shown
that quasi-steady nuclear burning models can produce both supersoft and
quasisoft radiation, but the model is too soft and/or luminous to fit
the r1-25 spectra.  

We note that there are other state-changing sources in external
galaxies.\footnote{see http://www.cfa.harvard.edu/$\sim$jfliu/ for a
  list of state changing sources} For example, there are nine state
changing sources in nearby spiral galaxy M33. If we study a large
enough sample of state-changers, we are likely to find examples of all
three (white dwarf, neutron star, and black hole) models. Continued
monitoring of these sources will play an important 
role in testing these models. It is also important to identify QSSs and
state-changers in the Magellanic Clouds and Galaxy, where many test of
the nature of the accretors can be conducted.

\acknowledgements

We would like to thank the {\it Swift} team for approving our ToO
request (Target ID: 35336). This research
has made use of the NASA/IPAC Extragalactic Database
(NED) which is operated by the Jet Propulsion Laboratory,
Caltech, under contract with the National Aeronautics and Space
Administration. This research was supported by HST Grant
AR-10948.01-A-0 and the Smithsonian Institution IR \& D award. BP is supported at Rutgers University in part by NSF award AST-0847157. We would like to thank the anonymous referee for comments that have helped to improve the paper.

\clearpage

\begin{figure}
\begin{center}
\includegraphics[angle=0,width=5in,height=4.5in]{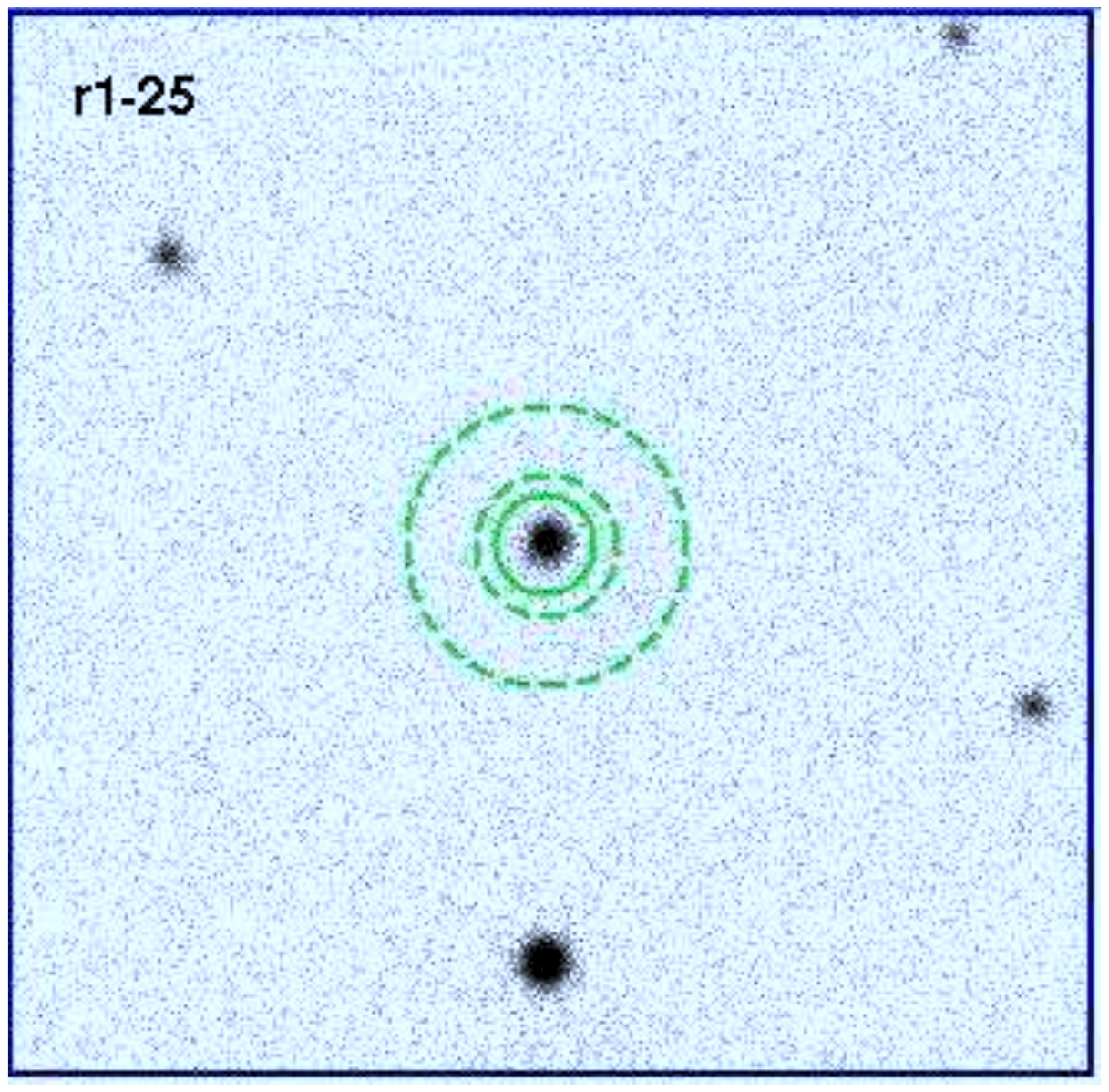}

\caption{ Source (solid) and background (dashed) apertures for r1-25,
  displayed on a $\sim$432 kilosecond merged Chandra HRC-I observation of M31.}
\end{center}
\end{figure}

\begin{figure}
\begin{center}
\includegraphics[angle=0,width=6in]{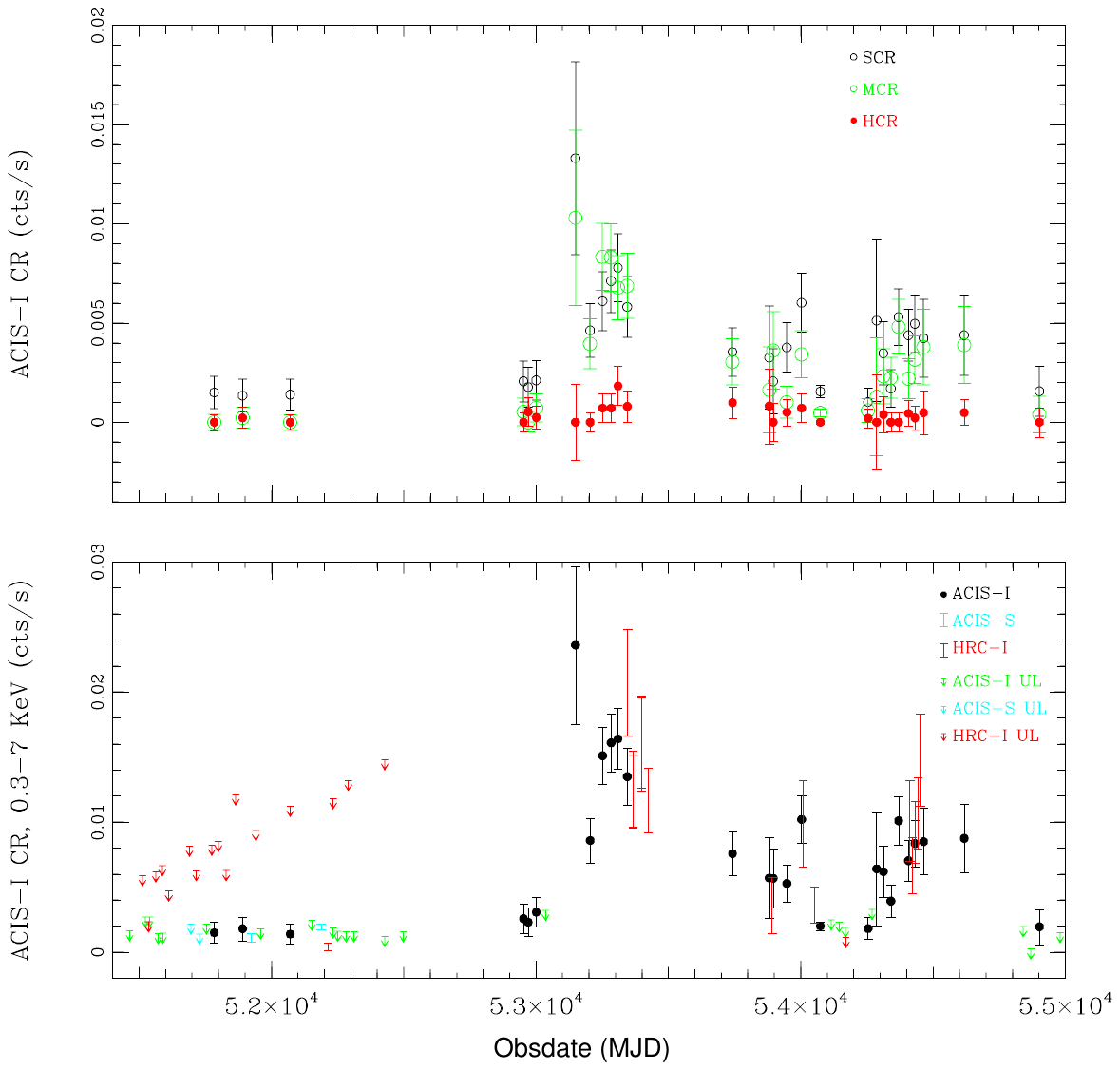}

\caption{ Light curve for r1-25. {\it Top Panel:} Count Rate in the
  Soft (0.3-1.1 keV), 
  Medium (1.1-2 keV), and Hard (2-7 keV) bands vs. Obsdate. {\it Bottom Panel}: Total count rate in ACIS-I units (in the energy
  range of 0.3-7 keV) vs. Obsdate. See text for how we converted ACIS-S and HRC-I observations to ACIS-I units.} 
\end{center}
\end{figure}      

\begin{figure}
\begin{center}
\includegraphics[width=2in]{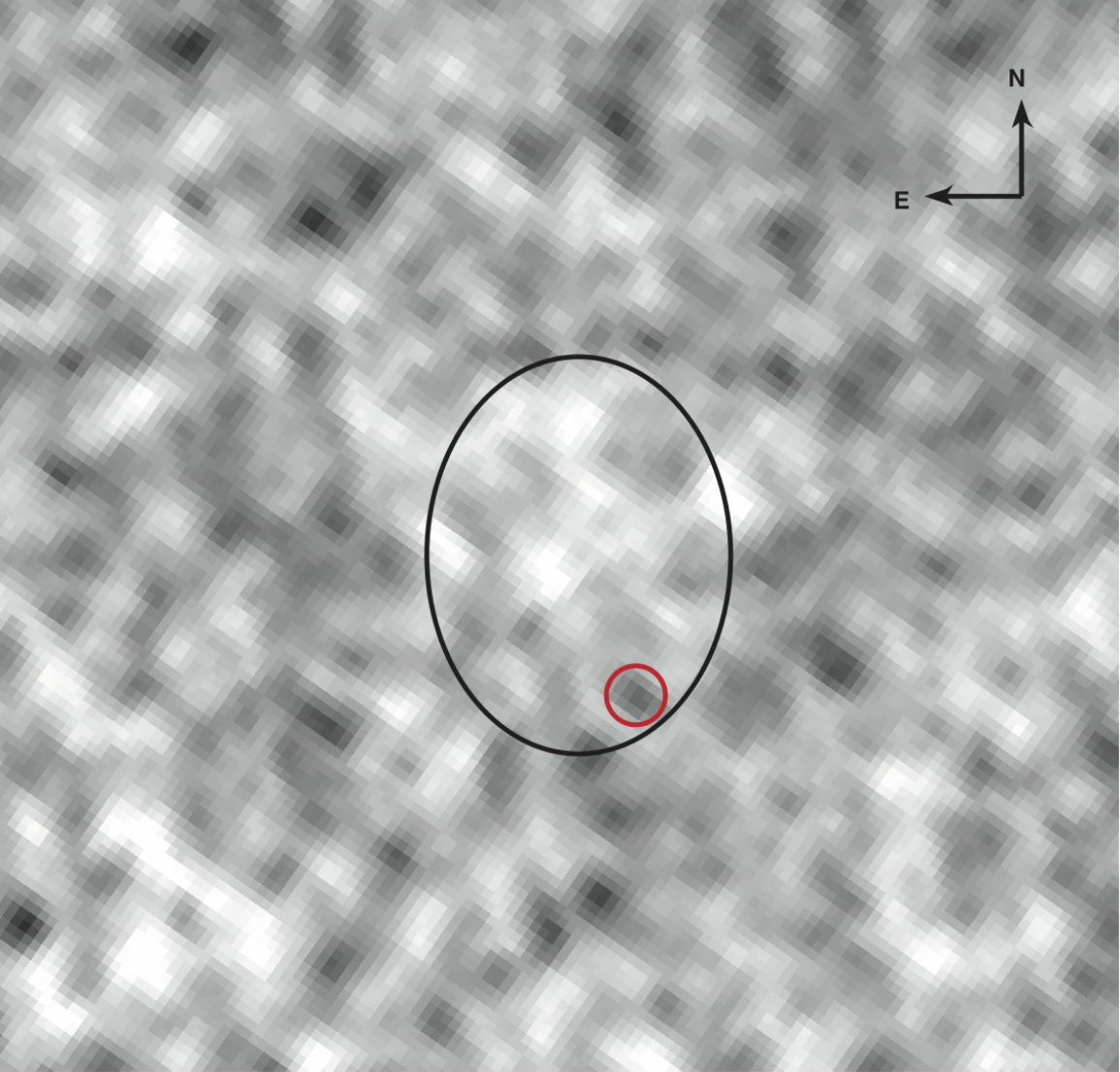}
\includegraphics[width=2in]{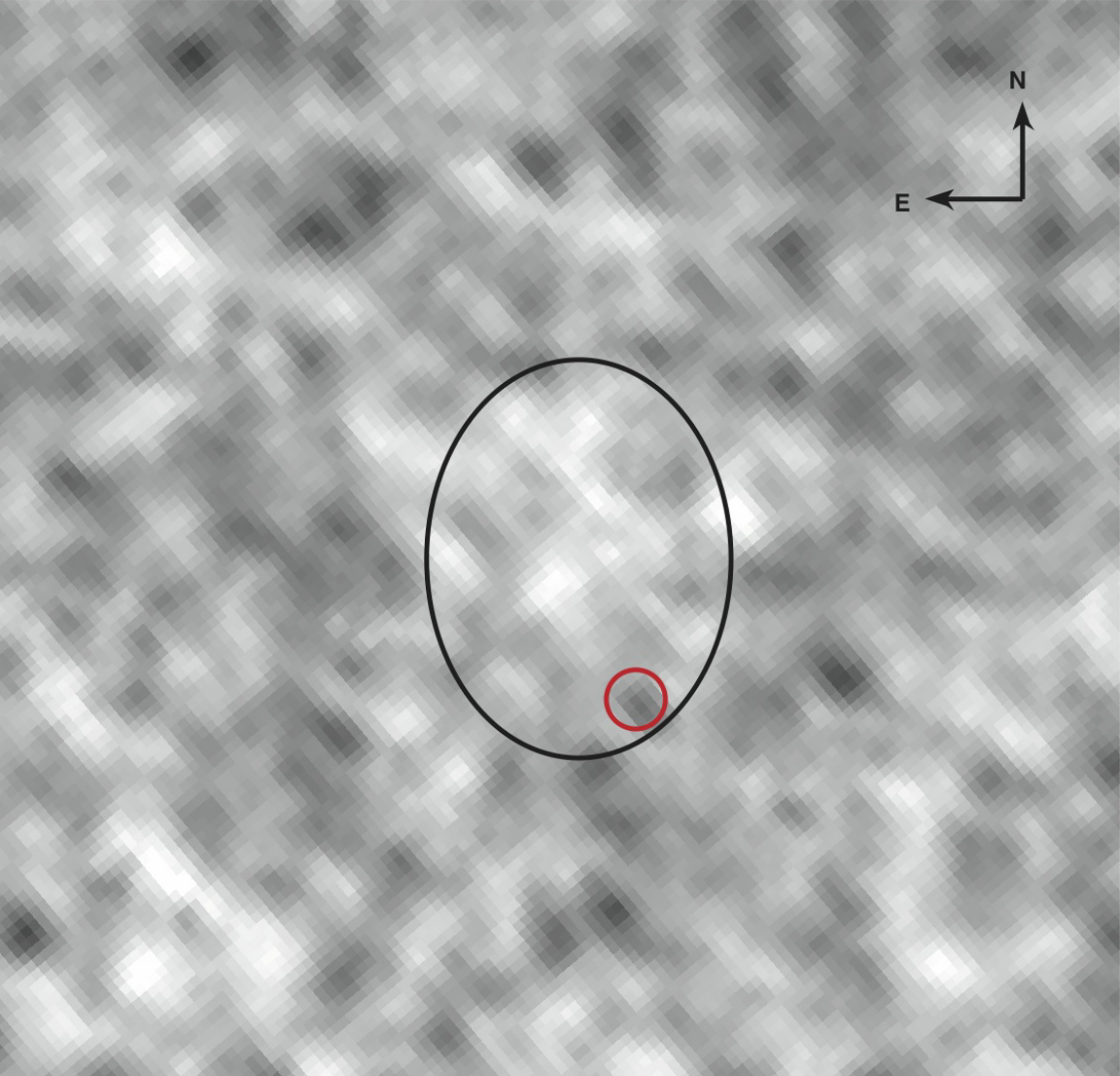}
\end{center}

\begin{center}
\includegraphics[width=2in]{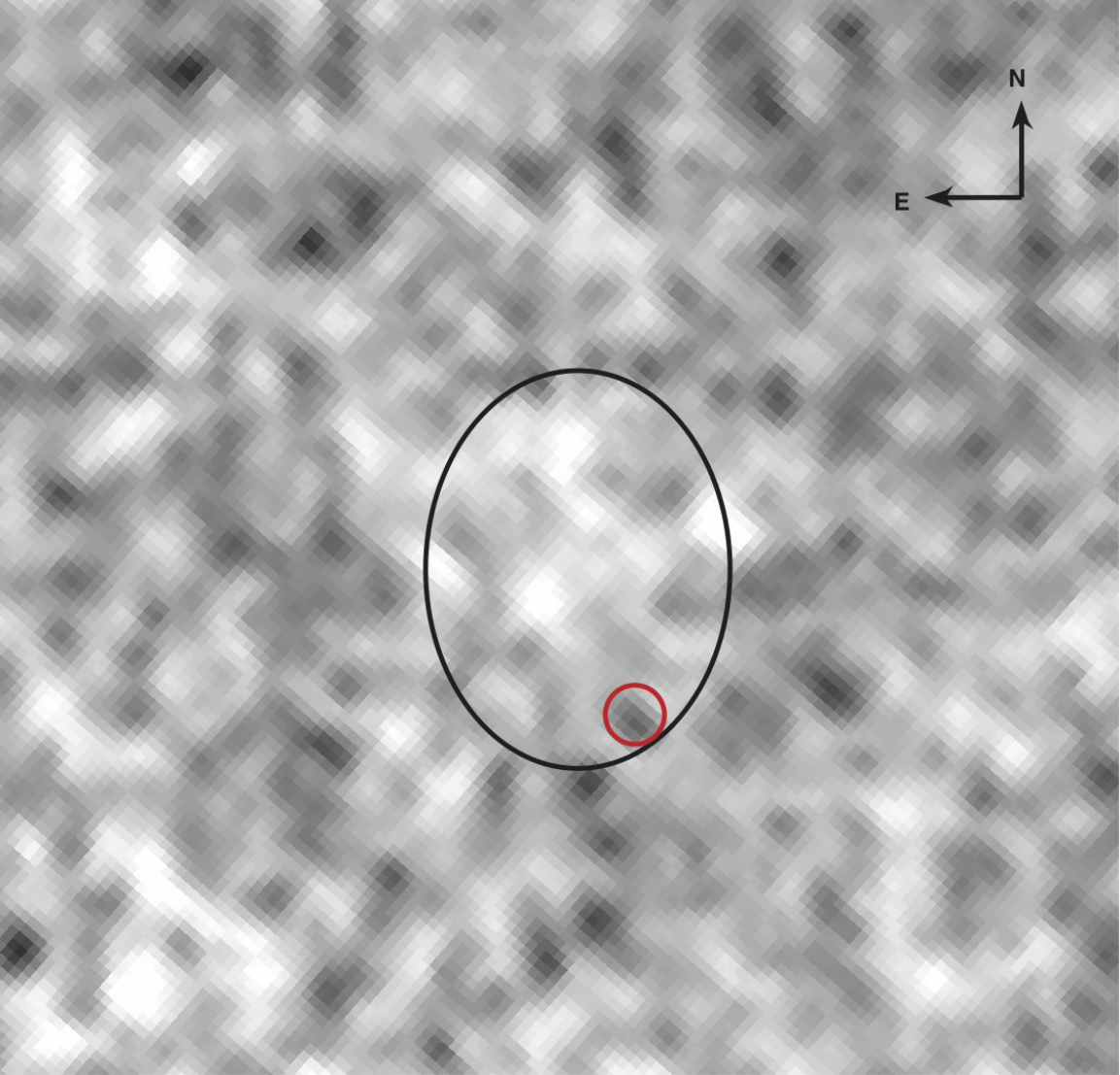}
\includegraphics[width=2in]{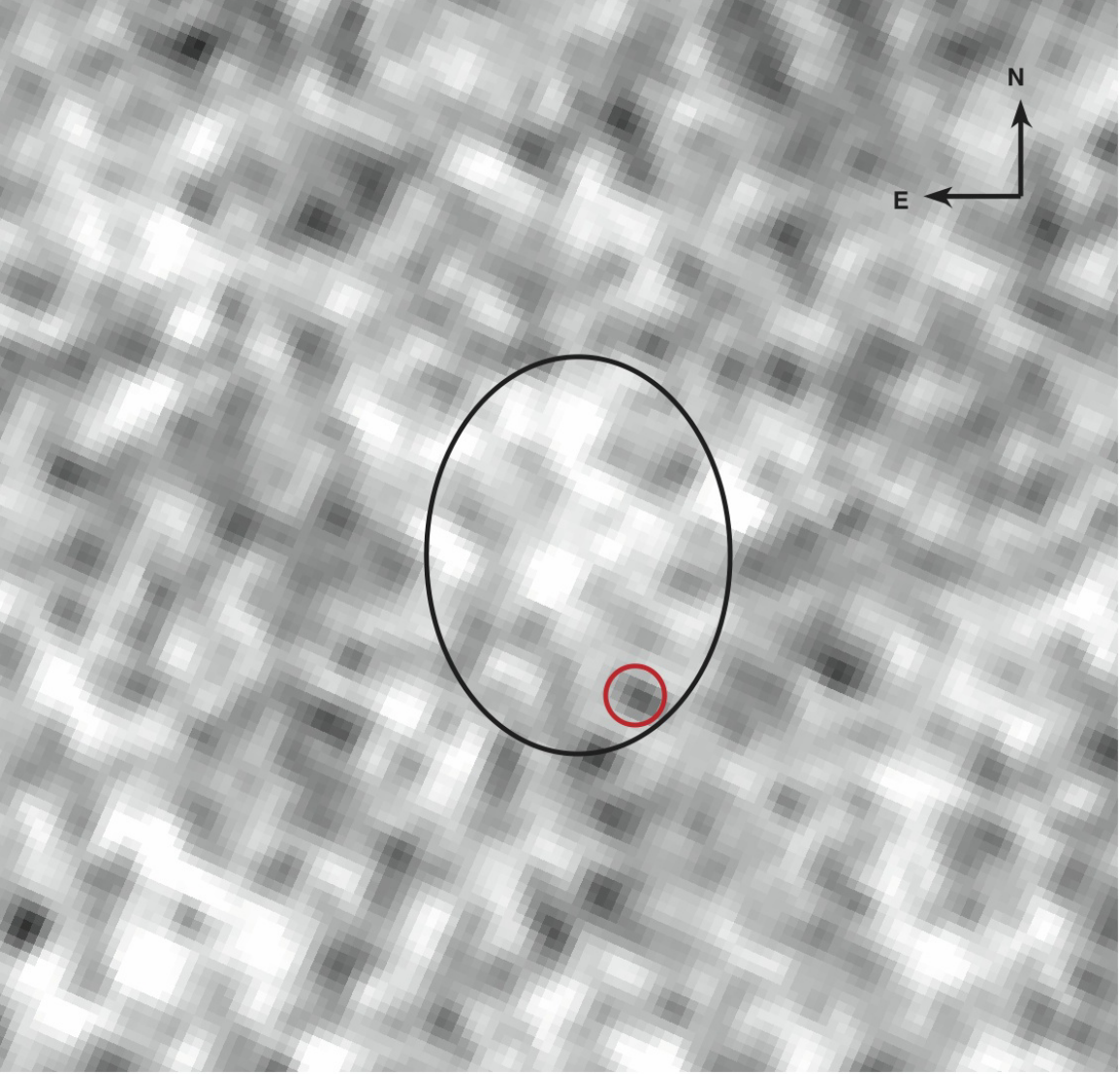}
\caption{HST ACS images of the location of r1-25, obtained in the
  F435W filter on four separate dates. {\it Top Left:}
  Observation j8vp03010, {\it Top Right:} observation j8vp05010, {\it
  Bottom Left:} Observation j9ju01010, {\it Bottom Right:} Observation
  j9ju06010. Each image is
  3$\arcsec$ on a side.  The black ellipse denotes the 3$\sigma$ X-ray
  positional error ellipse of r1-25. The ellipse has a semi-major axis
  (3$\sigma$ RA error) = 0.39$\arcsec$ and a semi-minor axis
  (3$\sigma$ DEC error) = 0.51 $\arcsec$. The red circle shows the
  only consistently 
  detected, resolved source found by DAOphot.       
}
\end{center}
\end{figure}

\begin{figure}
\begin{center}
\includegraphics[angle=0,width=5in,height=5in] {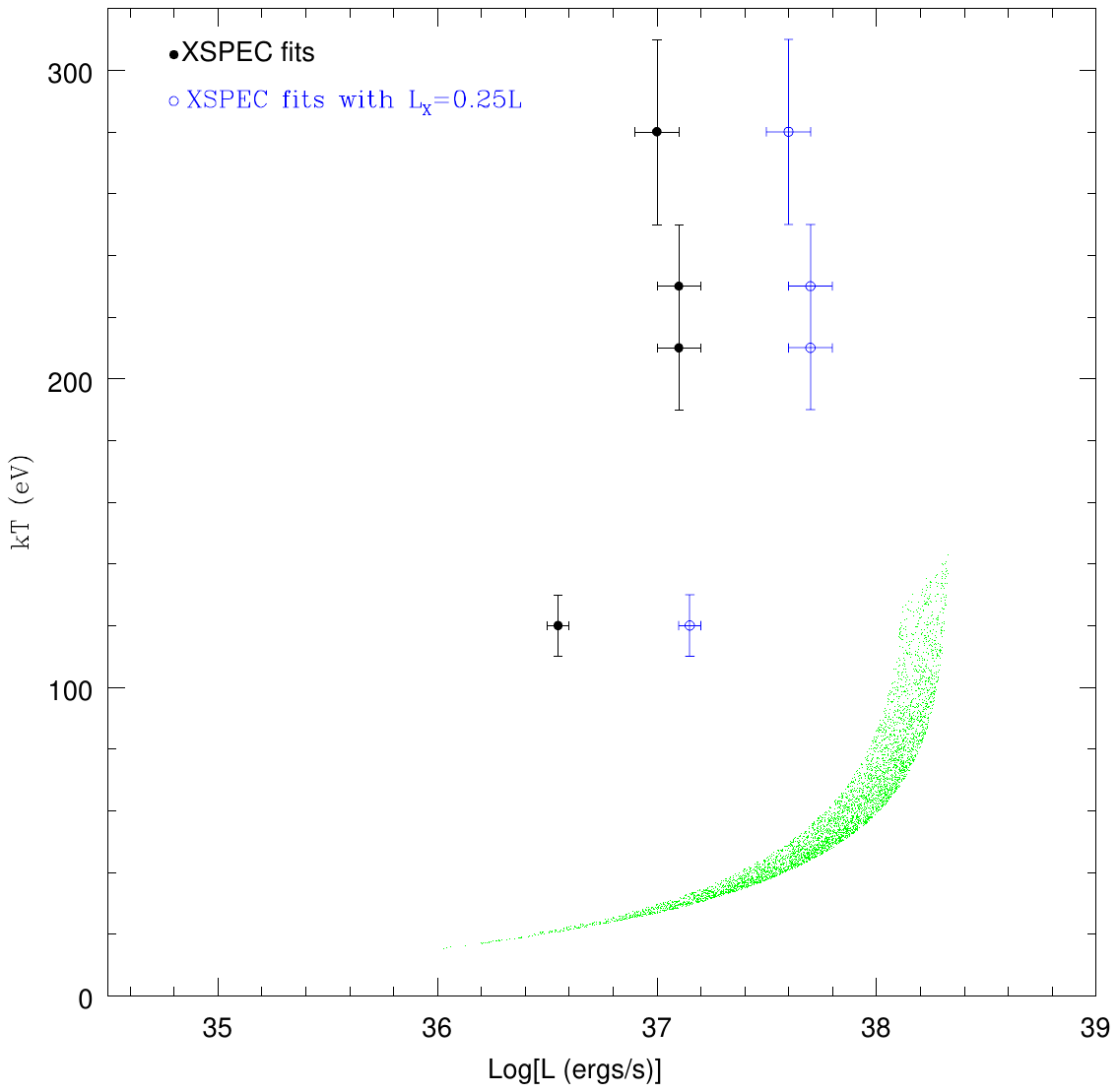}
\caption{Plot of $k\, T$
  vs. the logarithm of the bolometric luminosity, 
  LOG[L]. The black and blue points show the observations with XSPEC fits from Table 2 (assuming {\it N$_H$} =
1 $\times$ $10^{21}$ cm$^{-2}$). The solid
  black points represent XSPEC fits assuming L$_X$ = L, and the open
  blue points assume L$_X$ = $0.25L$. The green points represent the $k\, T$ and LOG[L]
  for various quasi-steady nuclear burning white dwarfs.    
}
\end{center}
\end{figure}

\begin{figure}
\begin{center}
\includegraphics[angle=0,width=5in,height=5in]{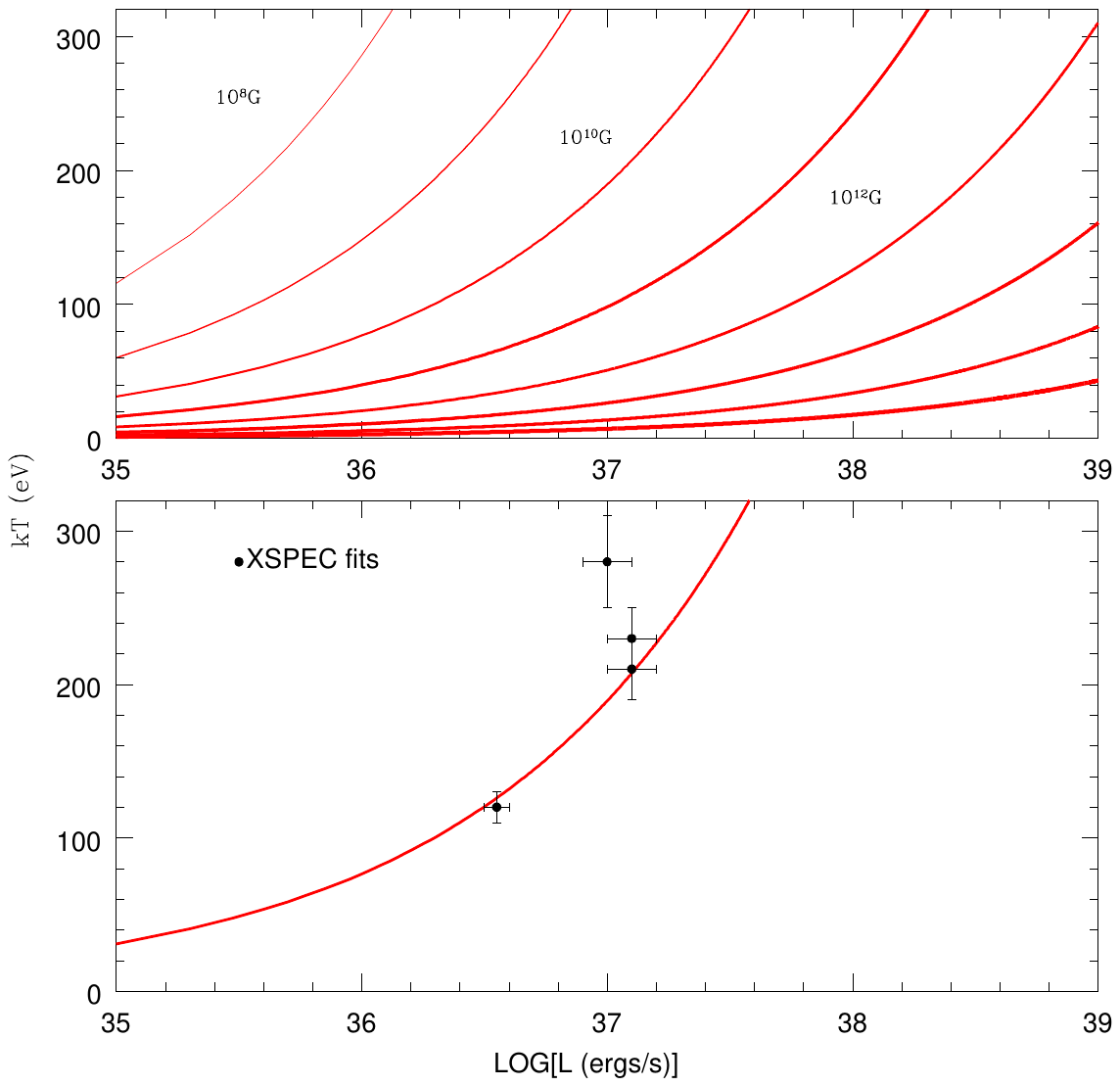}
\caption{ Plot of $k\, T$ versus $L$ for the neutron star model
  described by Equation 1.
{\it Top Panel:} Each curve corresponds to a fixed value of the
  magnetic field on the surface of the neutron star; $B_s$ changes by
  a factor of ten between curves. Since the field is not 
expected to vary this much over short times, the system should
evolve {\it along} the curves. Thus, if the luminosity increases or
decreases, so does the temperature. {\it Bottom Panel:} The
$B_s=10^{10}$~G curve is plotted with the r1-25 spectra. The black points represent XSPEC fits from Table 2 (assuming {\it N$_H$} =
1 $\times$ $10^{21}$ cm$^{-2}$). 
}
\end{center}
\end{figure}

\begin{figure}
\begin{center}
\includegraphics[angle=0,width=5in,height=5in]{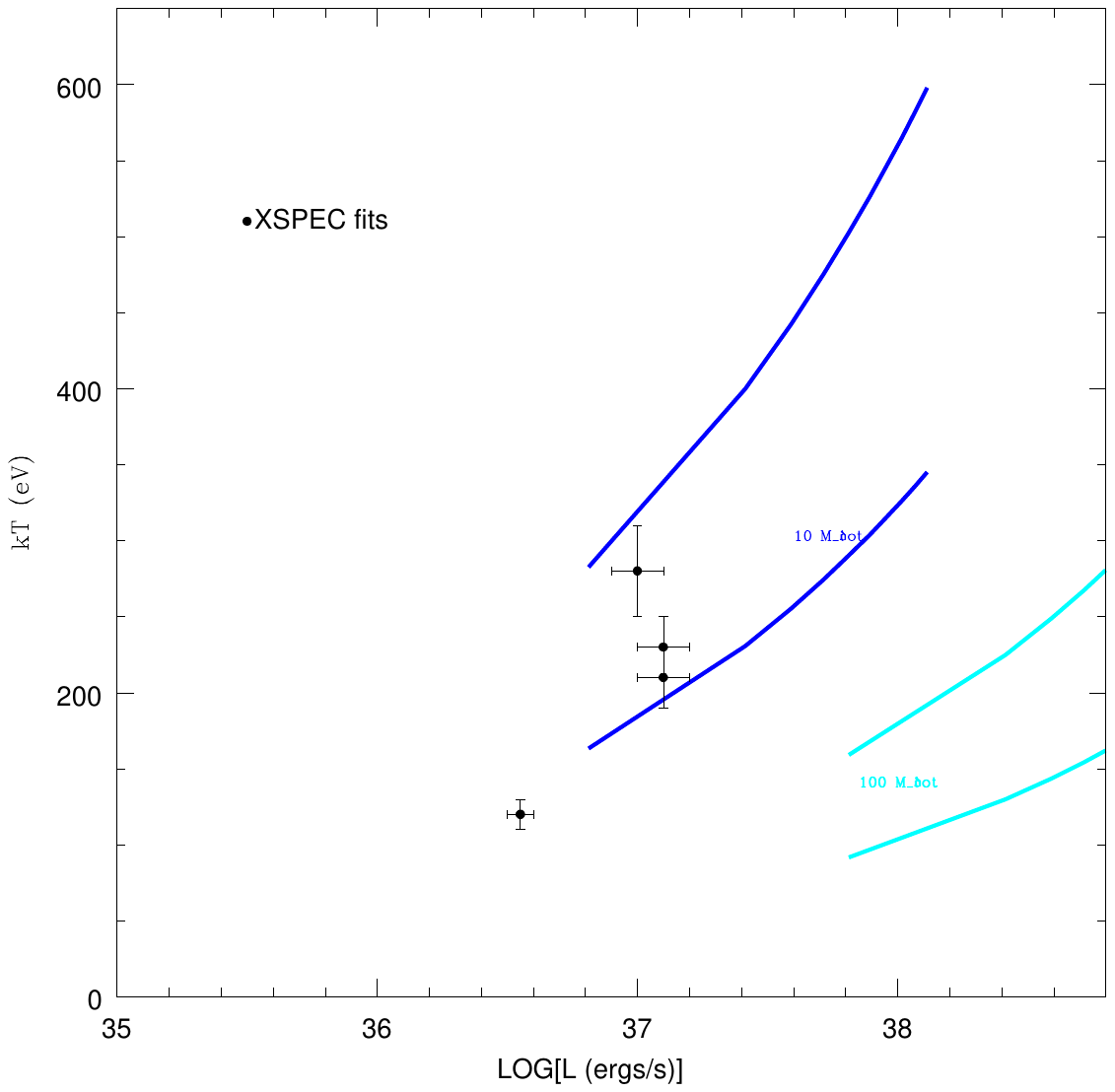}
\caption{Figure taken from \citet{Dis2010b}. It is a plot of
  $k\, T$ versus LOG[L] for the inner portion of the 
  accretion disk around black holes. Each pair of two
  curves of a single color corresponds to a fixed black hole mass
  which labels the regions between the curves. The upper curve of each
  color corresponds to a disk with inner radius $6 M_{BH} G/c^2$ (3r$_s$),
  while the lower curve corresponds to an inner disk with 9r$_s$. The point at the bottom (top) of each curve corresponds
  to the luminosity of the inner disk being $1\% L_{Eddington}$ ($10\%
  L_{Eddington}$). The curves are plotted with the r1-25 spectra. The black points represent the XSPEC fits from Table 2 (assuming {\it N$_H$} =
1.1 $\times$ $10^{21}$ cm$^{-2}$) 
}
\label{label1}
\end{center}
\end{figure}

\begin{deluxetable}{cccccccc}
\tablecolumns{8}
\setlength{\tabcolsep}{0.2in}
\tabletypesize{\scriptsize}
\tablecaption{Photometry for r1-25}
\tablehead{
\colhead{} & \colhead{} & \colhead{Exposure Time} & \colhead{SCR} & \colhead{MCR} & \colhead{HCR} & \colhead{TCR} & \colhead{Total}  \\ \colhead{ObsID} & \colhead{OBSMJD} & \colhead{seconds} & \colhead{$10^{-3}$s$^{-1}$} & \colhead{$10^{-3}$s$^{-1}$} & \colhead{$10^{-3}$s$^{-1}$} & \colhead{$10^{-3}$s$^{-1}$} & \colhead{Counts}   
}

\startdata
\cutinhead{ACIS-I Observations}
312		&	51783.8	&	4666		&	1.5 $\pm$ 0.8	&	0.0 $\pm$ 0.4	&	0.0 $\pm$ 0.4	&	1.5 $\pm$ 0.8	&	7 $\pm$ 4 \\
1581		&	51891.1	&	4404		&	1.3 $\pm$ 0.8	&	0.2 $\pm$ 0.5	&	0.2 $\pm$ 0.5	&	1.8 $\pm$ 0.9	&	8 $\pm$ 4	 \\
1583	 	&	52070.8	&	4903		&	1.4 $\pm$ 0.8	&	0.0 $\pm$ 0.4	&	0.0 $\pm$ 0.4	&	1.4 $\pm$ 0.8	&	7 $\pm$ 4	 \\
4678	 	&	52952.3	& 	3894		&	2.1 $\pm$ 1.0	&	0.5 $\pm$ 0.7	&	0.0 $\pm$ 0.5	&	2.6 $\pm$ 1.1	&	10 $\pm$ 4 \\
4679		&	52969.9	&	3820		&	1.8 $\pm$ 1.0	&	0.0 $\pm$ 0.5	&	0.5 $\pm$ 0.7	&	2.3 $\pm$ 1.1	&	9 $\pm$ 4 \\
4680	 	&	53000.3	&	4198		&	2.1 $\pm$ 1.0	&	0.7 $\pm$ 0.7	&	0.2 $\pm$ 0.6	&	3.1 $\pm$ 1.1	&	13 $\pm$ 5 \\
4682	 	&	53148.7	&	3945		&	13.3 $\pm$ 4.9	&	10.3 $\pm$ 4.4	&	0.0 $\pm$ 1.9	&	23.6 $\pm$ 6.1	&	23 $\pm$ 6 \\
4719	 	&	53203.9	&	4123		&	4.6 $\pm$ 1.4	&	3.9 $\pm$ 1.3	&	0.0 $\pm$ 0.5	&	8.6 $\pm$ 1.7	&	35 $\pm$ 7 \\
4720	 	&	53250.6	&	4108		&	6.1 $\pm$ 1.5	&	8.3 $\pm$ 1.7	&	0.7 $\pm$ 0.7	&	15.1 $\pm$ 2.2	&	62 $\pm$ 9 \\
4721 &	53282.9	&	4132		&	7.1 $\pm$ 1.6	&	8.3 $\pm$ 1.7	&	0.7 $\pm$ 0.7	&	16.1 $\pm$ 2.2	&	66 $\pm$ 9 \\
4722 &	53309.1	&	3894		&	7.8 $\pm$ 1.7	&	6.8 $\pm$ 1.6	&	1.8 $\pm$ 1.0	&	16.4 $\pm$ 2.3	&	63 $\pm$ 9 \\
4723	 &	53344.4	&	4035		&	5.8 $\pm$ 1.5	&	6.9 $\pm$ 1.6	&	0.8 $\pm$ 0.8	&	13.5 $\pm$ 2.2	&	51 $\pm$ 8 \\
7136		&	53741.8	&	3991		&	3.5 $\pm$ 1.2	&	3.0 $\pm$ 1.2	&	1.0 $\pm$ 0.8	&	7.6 $\pm$ 1.7	&	30 $\pm$ 7 \\
7137	 	&	53881.2	&	3952		&	3.3 $\pm$ 2.6	&	1.6 $\pm$ 2.2	&	0.8 $\pm$ 1.9	&	5.7 $\pm$ 3.1	&	7 $\pm$ 4	  \\
7138		 &	53895.7	&	4107		&	2.1 $\pm$ 1.6	&	3.6 $\pm$ 1.9	&	0.0 $\pm$ 1.0	&	5.7 $\pm$ 2.3	&	11 $\pm$ 4 \\
7139		 &	53947.0	&	3987		&	3.8 $\pm$ 1.2	&	1.0 $\pm$ 0.8	&	0.5 $\pm$ 0.7	&	5.3 $\pm$ 1.4	&	21 $\pm$ 6 \\
7140		&	54002.8	&	4117		&	6.0 $\pm$ 1.5	&	3.4 $\pm$ 1.2	&	0.7 $\pm$ 0.7	&	10.2 $\pm$ 1.9	&	42 $\pm$ 8 \\
7064 	&	54073.9	&	23238	&	1.5 $\pm$ 0.3	&	0.5 $\pm$ 0.2	&	0.0 $\pm$ 0.1	&	2.0 $\pm$ 0.3	&	46 $\pm$ 8 \\
7068	 	&	54253.9	&	7693		&	1.0 $\pm$ 0.7	&	0.6 $\pm$ 0.6	&	0.2 $\pm$ 0.5	&	1.8 $\pm$ 0.8	&	9 $\pm$ 4  \\
8192	 	&	54286.5	&	4073		&	5.1 $\pm$ 4.1	&	1.28 $\pm$ 3.0	&	0.0 $\pm$ 2.4	&	6.4 $\pm$ 4.4	&	5 $\pm$ 3 \\
8193	 	&	54312.1	&	4129		&	3.5 $\pm$ 1.6	&	2.3 $\pm$ 1.4	&	0.4 $\pm$ 0.9	&	6.2 $\pm$ 2.0	&	16 $\pm$ 5 \\
8194	 	&	54340.5	&	4033		&	1.7 $\pm$ 0.9	&	2.2 $\pm$ 1.0	&	0.0 $\pm$ 0.5	&	3.9 $\pm$ 1.3	&	16 $\pm$ 5 \\
8195	 	&	54369.6	&	3965		&	5.3 $\pm$ 1.4	&	4.8 $\pm$ 1.4	&	0.0 $\pm$ 0.5	&	10.1 $\pm$ 1.9	&	40 $\pm$ 7 \\
8186 	&	54407.2	&	4134		&	4.4 $\pm$ 1.3	&	2.2 $\pm$ 1.0	&	0.5 $\pm$ 0.7	&	7.0 $\pm$ 1.6	&	29 $\pm$ 6 \\
8187		&	54431.2	&	3837		&	5.0 $\pm$ 1.4	&	3.2 $\pm$ 1.2	&	0.2 $\pm$ 0.6	&	8.3 $\pm$ 1.8	&	32 $\pm$ 7 \\
9520	 	&	54463.7	&	3962		&	4.2 $\pm$ 2.0	&	3.8 $\pm$ 1.9	&	0.5 $\pm$ 1.1	&	8.5 $\pm$ 2.6	&	18 $\pm$ 5 \\
9529	 	&	54617.5	&	4113		&	4.4 $\pm$ 2.0	&	3.9 $\pm$ 1.9	&	0.5 $\pm$ 0.7	&	8.8 $\pm$ 2.6	&	18 $\pm$ 5 \\
10553	&	54901.6	&	4104		&	1.6 $\pm$ 1.3	&	0.4 $\pm$ 0.9	&	0.0 $\pm$ 0.7	&	2.0 $\pm$ 1.3	&	5 $\pm$ 3 \\
\cutinhead{ACIS-S Observations}
1854	 	&	51922.4	&	4694		&	4.3 $\pm$ 1.2	&	0.7 $\pm$ 0.6	&	0.0 $\pm$ 0.4	&	4.9 $\pm$ 1.3	&	23 $\pm$ 6 \\
1575 
	 	&	52187.0	&	37664	&	4.6 $\pm$ 0.4	&	0.4 $\pm$ 0.1	&	0.0 $\pm$ 0.1	&	4.9 $\pm$ 0.4	&	184 $\pm$ 15 \\
\cutinhead{HRC-I Observations}
1912	 	&	52214.0	&	46732	&	\nodata		&	\nodata		&	\nodata		&	0.5 $\pm$ 0.1	&	21 $\pm$ 6\\
5925	 	&	53345.8	&	46311	&	\nodata		&	\nodata		&	\nodata		&	14.6 $\pm$ 0.6	&	678 $\pm$ 28\\
6177		 &	53366.3	&	20038	&	\nodata		&	\nodata		&	\nodata		&	8.8 $\pm$ 0.7	&	176 $\pm$ 13\\
5926	 	&	53366.8	&	28268	& 	\nodata		&	\nodata		&	\nodata		&	8.7 $\pm$ 0.6	&	246 $\pm$ 16\\
6202	 	&	53398.1	&	18046	&	\nodata		&	\nodata		&	\nodata		&	11.3 $\pm$ 0.8	&	204 $\pm$ 14\\
5927	 	&	53398.8	&	27000	&	\nodata		&	\nodata		&	\nodata		&	11.4 $\pm$ 0.7	&	308 $\pm$ 18\\
5928	 	&	53422.7	&	44856	&	\nodata		&	\nodata		&	\nodata		&	8.2 $\pm$ 0.4	&	369 $\pm$ 20 \\
7283	 	&	53891.3	&	19942	&	\nodata		&	\nodata		&	\nodata		&	3.4 $\pm$ 0.4	&	67 $\pm$ 9 \\
7284	 	&	54008.9	&	20002	&	\nodata		&	\nodata		&	\nodata		&	10.6 $\pm$ 0.7	&	212 $\pm$ 14 \\
7285	 	&	54052.3 	&	18517	&	\nodata		&	\nodata		&	\nodata		&	3.8 $\pm$ 0.5	&	70 $\pm$ 9 \\
8526	 	&	54411.6	&	19944	&	\nodata		&	\nodata		&	\nodata		&	9.6 $\pm$ 0.7	&	191 $\pm$ 14  \\
8527	 	&	54421.8	&	19981	&	\nodata		&	\nodata		&	\nodata		&	6.3 $\pm$ 0.6	&	126 $\pm$ 12 \\
8528	 	&	54432.8	&	19975	&	\nodata		&	\nodata		&	\nodata		&	7.7 $\pm$ 0.6 	&	154 $\pm$  13 \\
8529	 	&	54441.6	&	18923	&	\nodata		&	\nodata		&	\nodata		&	8.9 $\pm$ 0.7	&	168 $\pm$ 13 \\
8530	 	&	54451.5	&	19915	&	\nodata		&	\nodata		&	\nodata		&	12.4 $\pm$ 0.8	&	247 $\pm$ 16 \\	
\enddata
\tablecomments{Columns 4, 5, and 6 are the count rates in the Soft (0.3-1.1
  keV), Medium (1.1-2 keV), Hard (2-7 keV) bands, respectively. All rates are corrected by a vignetting factor.}
 
\end{deluxetable}

\clearpage

\begin{deluxetable}{cccccccc}
\tabletypesize{\scriptsize}
\tablecaption{Spectral Fitting Results}
\tablehead{
\colhead{} & \multicolumn{3}{c}{{\it N$_H$} = 1.1 $\times$ $10^{21}$ cm$^{-2}$} &
\colhead{} & \multicolumn{3}{c}{{\it N$_H$} = 6.4 $\times$ $10^{21}$ cm$^{-2}$} \\
\cline{3-5} \cline{7-8} \\
\colhead{Obs ID} & \colhead{$k\, T$ (keV)} & \colhead{Normalization (10$^{-6}$)} & \colhead{L$_{X}$ (10$^{37}$ ergs s$^{-1}$)} & \colhead{} & \colhead{$k\, T$ (keV)} & \colhead{Normalization (10$^{-6}$)} &
\colhead{L$_{X}$ (10$^{37}$ ergs s$^{-1}$)}
}
\startdata
     1575 & 0.12 $\pm$ 0.01 & 0.73$^{+0.09}_{-0.08}$ & 0.40 & & 0.065 $\pm$ 0.003 & 160$^{+50}_{-40}$ & 97\\
     4720 & 0.28 $\pm$ 0.03 & 1.7 $\pm$ 0.3 & 1.0 & & 0.19 $\pm$ 0.02 & 8$^{+3}_{-2}$ & 4.8\\
     4721 & 0.23 $\pm$ 0.02 & 2.0$^{+0.4}_{-0.3}$ & 1.2 & & 0.17 $\pm$ 0.01 & 10$^{+4}_{-3}$ & 6.0\\
     4722 & 0.21 $\pm$ 0.02 & 2.4$^{+0.5}_{-0.4}$ & 1.4 & & 0.11 $\pm$ 0.01  & 60$^{+40}_{-20}$ & 36 \\
\enddata

\tablecomments{Spectral fits were carried out in the energy range
  0.3-8.0 keV, using a simple absorbed blackbody model (wabs model was used for absorption). Luminosities are given for the same energy range and
  are corrected for assumed absorption. All uncertainties are
  1$\sigma$ in size.}               
\end{deluxetable}

\clearpage

\end{document}